\documentclass{JHEP3} 

\usepackage{epsfig,multicol}

\title{Gravitational duality near de Sitter space}

\author{Bernard Julia\thanks{Laboratoire de Physique th\' eorique  UMR 
8549 ENS/CNRS.}~, J\' er\^ ome Levie and S\' ebastien Ray \\
         LPT, 24 rue Lhomond, 75231 Paris CEDEX 05, FRANCE\\
        E: \email{bjulia levie sray AT lpt.ens.fr}}

\accepted{July 26, 2005 . LPT/ENS-preprint 05/23}                

\preprint{\hepth{0507262}} 

\abstract{Nonlinear gravitational instantons ``$\Lambda$-instantons''  are defined here 
for any given value $\Lambda$ of the cosmological constant. A multiple of the 
Euler characteristic appears as an upper bound for the four dimensional de Sitter action 
and as a lower bound for a family of quadratic actions. Our new self-duality equation
for the MacDowell Mansouri tensor implies the full nonlinear Einstein's equations
for a given value of the cosmological constant. The de Sitter action itself
is found to be equivalent to a simple and natural quadratic action. 
In this paper we also describe explicitly the 
reparameterization and duality invariances of gravity (in 4 dimensions) 
linearized about de Sitter space. A noncovariant doubling of the 
fields using the Hamiltonian formalism leads to first order time evolution
with manifest duality symmetry. As a special case  we recover the linear flat 
space result of Henneaux and Teitelboim by a smooth limiting process.} 


\begin{document} 

\section{Introduction}

We uncover here two new properties of General Relativity. The existence of
duality symmetries which preserve the equations of motion and sometimes the
action are important features of a theory as they
frequently allow us to relate a strong coupling regime to a
perturbative one which is more amenable to computation. In a linearized 
theory around flat
space it is generally true that one can (Hodge) dualize differential form
fields to get similar and sometimes even identical equations after
dualization. It is much more exceptional to have nonlinear duality
symmetries \cite{Jprogress}. Their applications from Statistical mechanics 
to Electromagnetism or String theory are well known.  
We shall present a step towards nonlinear duality symmetries,
namely the existence of a duality rotation group for linearized gravity
around de Sitter space rather than around flat space. We recall the 
surprising symmetry between General Relativity and its dual theory in 
the linear approximation around Minkowski space \cite{HT}.
Now in the quest for duality symmetries one may consider starting 
by considering self-dual solutions. In fact we shall present a new 
instanton 
equation for gravity at the full nonlinear level as a motivation for our duality 
analysis, the same deformation of the Riemannian curvature tensor will 
appear there as in the duality study.  

The first result we discuss is that
one can define a one parameter family of nonlinear gravitational instantons 
parameterized precisely by the value of the cosmological constant.  Before 
defining this new self-duality (instanton) condition corresponding to any 
given value of the cosmological constant we shall review in section 3 
the three previously known types of gravitational instantons. The first two types 
correspond respectively to special Ricci flat solutions or  Einstein spaces
for arbitrary values  of the cosmological constant. The third type is
slightly different, implying only the self-duality of the Weyl (conformal) 
tensor rather than that of the full Riemann curvature tensor. Our new 
"$\Lambda$ instantons" reduce to the first type of (ordinary) instantons 
when one sets the cosmological constant $ \Lambda$ to zero.
 
To obtain the second 
result of this paper we shall generalize the ordinary gravitational instantons 
and verify the duality invariance of the linearization of 
General Relativity around de Sitter space. This boosts our confidence 
in good duality 
properties of gravity at the nonlinear level  without assuming any isometry 
\cite{JU}. In section 5 we introduce prepotentials and in section 6 their 
duality transformation
laws.  Some of these correspond to diffeomorphisms whereas others are
(pre)gauge  symmetries. Finally section 7 is devoted to check the invariance of
the action.   

\section{Gravity equations with a cosmological constant}

Let us consider four dimensional spacetime with a metric either of Lorentzian 
signature (-,+,+,+) or of euclidean (or (2+, 2-)) signature for the 
existence of instantons. Our main conventions are those of the textbook by 
Hawking and Ellis. 
The cosmological constant is denoted by $\Lambda$ and we focus on the
nonvanishing case. For definiteness the  computations starting with section 4 
will actually be done for a positive value of $\Lambda$  and Lorentzian 
signature.
The usual equations for the Einstein tensor read 
$$G_{\mu\nu}=-\Lambda g_{\mu\nu}.$$ 
Flat space can be seen either as a gravitational instanton or anti-instanton. 
There are (anti)-self-dual tensors in (euclidean) de Sitter space as well as 
in the flat space case, however they are not the chiral parts of the Riemann 
curvature tensor, they are rather the chiral parts of the Weyl tensor 
$C_{\mu\nu\rho\sigma}$ 
which  vanishes there.
Let us recall its definition:
\begin{eqnarray}
C_{\mu\nu\rho\sigma} \equiv R_{\mu\nu\rho\sigma} 
+\frac{1}{2} (g_{\mu\sigma}R_{\rho\nu}-g_{\mu\rho}R_{\sigma\nu} 
 +g_{\nu\rho}R_{\sigma\mu}-g_{\nu\sigma}R_{\rho\mu} )
+\frac{1}{6}(g_{\mu\rho}g_{\nu\sigma} - g_{\mu\sigma}g_{\nu\rho})R\
\label{weyl}
\end{eqnarray}
Actually on shell the Weyl tensor $C_{\mu\nu\rho\sigma}$ is equal to a new 
tensor $Z_{\mu\nu\rho\sigma}$ defined as 
follows:
\begin{eqnarray}
Z_{\mu\nu\rho\sigma} \equiv R_{\mu\nu\rho\sigma} - 
\frac{\Lambda}{3}(g_{\mu\rho}g_{\nu\sigma} - g_{\mu\sigma}g_{\nu\rho})\,.
\label{Z}
\end{eqnarray}
How does one discover this tensor 
$Z_{\mu\nu\rho\sigma}$? 
One way to find it  is to require the form of the
equations of motion and of the torsion Bianchi identity to be the same. This
is well known in the absence of cosmological constant and it turns out that
the only required modification to accommodate a nonvanishing $\Lambda$ is
precisely to replace $R_{\mu\nu\rho\sigma}$ by $Z_{\mu\nu\rho\sigma}.$ 
Symmetry properties and even topological
inequalities can be extended from one to the other to a very large extent. 
In fact $Z_{\mu\nu\rho\sigma}$  
obeys simple translation and also rotation Bianchi identities
\begin{eqnarray}
Z_{\mu[\nu\rho\sigma]}\equiv 0\\
\nabla_{[\mu}Z_{\nu\rho]\sigma\tau}\equiv 0
\label{Cbianc}
\end{eqnarray}
(there is no torsion.)
 
\noindent The equally beautiful equations of motion read:
\begin{equation}
Z_{\mu\nu} \equiv {Z_{\mu\rho\nu}}^\rho = 0.
\label{Eq}
\end{equation}
(Note that this would be an identity for the Weyl tensor).
\noindent Let us remark that the dual tensor:
\begin{equation}
\tilde Z_{\mu\nu\rho\sigma} \equiv \frac{1}{2} 
\epsilon_{\mu\nu\alpha\beta} {Z^{\alpha\beta}}_{\rho\sigma}
\label{Cdual}
\end{equation}
\noindent obeys the same equations but with reverse meaning. The torsion
Bianchi identity becomes the dual equation of motion $\tilde Z_{\mu\nu}=0$, 
and the equation of motion has the form of the first Bianchi identity
$\tilde Z_{\mu[\nu\rho\sigma]}=0  $ for the dual $\tilde Z_{\mu\nu\rho\sigma}$
tensor.

\section{$\Lambda$-instantons}

From the above remarks it is natural  to define the 
(anti)-$\Lambda$-instantons 
with given value $\Lambda$ of the cosmological constant by the equation 
\begin{equation}
\tilde Z_{\mu\nu\rho\sigma} = \pm  Z_{\mu\nu\rho\sigma}
\label{LInst}
\end{equation}
They automatically obey the second order Einstein equations for that particular
value of $\Lambda.$ We recall that similar equations define the usual 
gravitational (anti)-instantons. For Ricci flat metrics we further require that
\begin{equation}
*R_{\mu\nu\rho\sigma} = \pm  R_{\mu\nu\rho\sigma}
\label{Inst}
\end{equation}
where the star means Hodge dualisation on the second pair of indices. These
are nothing but ($\Lambda$=0)-instantons.

\noindent For Einstein spaces i.e. for an a priori undetermined value 
of the cosmological constant, one may speak of ``cosmoinstantons''  because
the Einstein equations can be equivalently rewritten
\begin{equation}
*\tilde R_{\mu\nu\rho\sigma} =   R_{\mu\nu\rho\sigma}
\label{SInst}
\end{equation}
where the tilde means dualization on the first pair of indices as above \cite{BB,EF}.
However we could reject the name instanton in that cas because the equations remain 
of second order.
There is a Hitchin-Bogomolny-Prasad-Sommerfield type inequality bounding below
the quadratic Yang action by multiples respectively of the 
Pontryagin and Euler invariants. Another similar inequality has been
introduced in \cite{DN} in relation with the  conformal instantons ie those instantons for 
which the Weyl tensor is self-dual (say on its first pair of indices). 
We may now notice that the ordinary  (anti)-instantons and more generally 
(anti)-$\Lambda$-instantons 
are instances of cosmoinstantons and are  also (anti)-conformal instantons.

$\Lambda$-instantons saturate analogously a BPS condition obtained by
replacing the curvature tensor $R$ by the new tensor $Z$ in the Yang action. 
The corresponding action $S_\Lambda^{(2)}\equiv  \int tr(Z.Z) d^4x$
is bounded below by $J_\Lambda(g)$ where we defined 
$J_\Lambda(g)\equiv\int tr(Z \tilde Z) d^4x.$
The right hand side lower bound of the BPS condition is still proportional 
to the signature invariant.

However the double self-duality equations 
$\tilde Z_{\mu\nu\rho\sigma} = * Z_{\mu\nu\rho\sigma}$ does not seem a 
priori 
interesting for our purpose as it only implies 
Einstein's equations for some arbitrary value of the cosmological constant,
exactly as in the case of ordinary duality $\tilde R _{\mu\nu\rho\sigma}= 
* R_{\mu\nu\rho\sigma}.$ 
Note however that this ``double self-duality'' leads to a BPS bound that is 
not a topological invariant, we find:
\begin{equation}
I_\Lambda(g)\equiv\int tr(\tilde Z \tilde Z) d^4x = 32\pi^2 \chi_{Euler}
-4/3 \,\Lambda  \, S_\Lambda
\label{LBPS}
\end{equation}
where
\begin{equation}
S_\Lambda = \int [R - 2\Lambda] \sqrt{g}d^4x
\end{equation}
is the Hilbert-Einstein-de Sitter action under consideration.
This simple computation shows that the quadratic action 
$I_\Lambda(g)$ is equivalent to the a priori
better quantum behaved de Sitter action. 

Equation \ref{LBPS} is a rederivation of the famous MacDowell Mansouri formula
\cite{MDM} for de Sitter action. Again we have
$$ S_\Lambda^{(2)} \ge I_\Lambda(g).$$
The bound given above is not anymore the Euler topological
invariant, it differs  from it by a multiple of $S_\Lambda.$ 
The expression $I_\Lambda(g)$ is non negative hence
\begin{equation}
\Lambda S_\Lambda  \le 24\pi^2 \chi_{Euler}. 
\end{equation}
We can rewrite the BPS inequality  as 
\begin{equation}
 \Lambda S_\Lambda \ge 24 \pi^2 \chi_{Euler} - 3/4 \int tr(Z.Z) d^4x
\label{SBOUND}
\end{equation}
or more suggestively
\begin{equation}
 \Lambda S_\Lambda +3/4S_\Lambda^{(2)}    \ge 24 \pi^2 \chi_{Euler}. 
\label{SBOUND'}
\end{equation}
We find that exactly for cosmoinstantons we have equality whatever the value
of the cosmological constant. We hope to return to this in the future.

\section{Equations near de Sitter geometry and linearized hamiltonian action}
 
  At the linear level one may replace the covariant derivative $\nabla$ by 
$\partial$ in the rotation Bianchi identity. This identity is the same as its 
dual because there is no torsion (even off shell) and hence 
$Z_{\mu\nu\rho\sigma}$ is symmetric 
under the exchange of its antisymmetric pairs of indices.
Let us rephrase the  result we have obtained: if one is careful to replace the 
Riemann
tensor $R_{\mu\nu\rho\sigma}$ by the $Z_{\mu\nu\rho\sigma}$ 
tensor which vanishes on de Sitter shell, the linearized 
equations exhibit self-duality 
near de Sitter space. This extends the result of Henneaux and Teitelboim
\cite{HT}  to the case of a non-vanishing cosmological constant. The idea 
of considering this generalization was also proposed in \cite{DS} and 
was motivated by supergravity.
We will now analyze in detail this symmetry and define dual potentials as
well as prepotentials and their duals to exhibit a continuous (one
parameter) symmetry group of the action.

We shall use an SO(3) covariant choice of coordinates covering one half of
de Sitter spacetime, the so-called planar coordinates: the $dS_4$ metric can 
be written 
\begin{equation}
ds^2=-dt^2+f^2(t)\delta_{ij}dx^i dx^j.
\end{equation}
Our latin indices refer to spatial directions whereas greek ones are four 
dimensional. Up to a constant we have
$f(t)=\exp\left(t\sqrt{\Lambda/3}\right)=\exp(Kt)$,
where it is convenient to define the inverse de Sitter radius $K\equiv
\sqrt{\Lambda/3}.$
\noindent Following \cite{HA} we can write the general gravity action 
in hamiltonian or 3+1  form namely 
\begin{equation}
S=\int d^4x \left[\pi^{ij}\dot{g}_{ij} + Ng_3^{1/2}(R^{(3)}-2\Lambda) + N 
g_3^{-1/2}\left(\frac{1}{2}\pi^2 - g_{ik}g_{jl}\pi^{ij}\pi^{kl} \right) + 
2N_i\left(\partial_j \pi^{ij}+\Gamma^{(3)i}_{jk}\pi^{jk}\right) \right]
\end{equation}
with
\begin{eqnarray*}
N=(-g^{00})^{-1/2},\;&N_i=g_{0i},\\
g_3=\det(g_{ij}),\; &\pi=g_{ij}\pi^{ij}.\\
\end{eqnarray*}
Here   $g_{ij}$ is the spatial metric, $\pi^{ij}$ its conjugate momentum
(density), $R^{(3)}$ is the 3D scalar curvature and 
$\Gamma^{(3)}$  the 3D Levi-Civita connection associated to $g_{ij}.$

\subsection{Linearization} 

Let us now expand around  de Sitter background. 

\begin{eqnarray*}
g_{ij}=f^2(t)\delta_{ij}+h_{ij},\; & \pi^{ij}=\bar\pi^{ij}+p^{ij},\\
N=1+n,\;& N_i=n_i.
\end{eqnarray*}
We find the background value of the  momenta: 
\begin{equation}
\bar\pi^{ij}=-2K f \delta^{ij}.
\end{equation}

Now to second order the various terms read (using the flat euclidean metric to
raise and lower the indices and to take traces (for instance $h$ and $p$)):

\begin{eqnarray}
g_3&=& f^6 \left(1+f^{-2}h+\frac{1}{2}f^{-4}(h^2-h^{ij}h_{ij}) \right)\\
-\sqrt{g_3} R^{(3)}  &=& f^{-1}( \Delta h-\partial_{i} \partial_{j} h^{ij}) + 
f^{-3} \left[ \frac{1}{2}h(\Delta h - \partial_i \partial_j h^{ij})
 \right.\nonumber\\
&&- h^{ij} (\partial_i \partial_j h + \Delta h_{ij} - 2\partial_i 
\partial^k h_{jk}) 
- \frac{3}{4}\partial^i h^{jk} \partial_i h_{jk} + \frac{1}{2} \partial^i 
h^{jk} \partial_j h_{ik} \nonumber\\
&& - \partial^i h_{ij} \partial^j h + \partial^i 
h_{ij} \partial_k h^{jk} +\left.\frac{1}{4}\partial_i h \partial^i h
 \right]\\
\frac{1}{2}\pi^2-g_{ik}g_{jl}\pi^{ij}\pi^{kl} &=& 
-f^2\left(f^{2}p^{ij}p_{ij} - 2K\, f p^{ij}h_{ij} 
+ 4K^2 h^{ij}h_{ij}\right)\nonumber\\ 
&&+\frac{f^2}{2}\left( f p - 2K\, h \right)^2
-2Kf^4(fp-2K\, h) + 6K^2f^6 
\end{eqnarray}
 It follows that the quadratic action near de Sitter space is
\begin{equation}
S=\int dx^4 \left(p^{ij}\dot h_{ij} - H - nC_1 - n_iC_2^i\right),
\end{equation}
\noindent with :
\begin{eqnarray}
 H&=&f^{-3}\left(\frac{1}{4}\partial^i h^{jk} \partial_i h_{jk} - 
\frac{1}{4}\partial^i h \partial_i h + \frac{1}{2} \partial^i h \partial^j 
h_{ij} - \frac{1}{2}\partial_i h^{ij} \partial^k h_{jk} \right) 
\nonumber\\
 && +f^{-1} (f^2p^{ij} p_{ij} - 2K\, fp^{ij} h_{ij} +K^2 h_{ij}h^{ij}) 
-\frac{1}{2} p(fp - 2K\, h)   ,\\
 C_1&=& - f^{-1} \left( \partial^i \partial^j h_{ij} - 
\Delta h\right) + 2f K(fp +  Kh)  ,\\
 C_2^i&=&-2\partial_j p^{ij} + 2f^{-1}K \left( 2\partial_j 
h^{ij} - \partial^i h \right) .
\end{eqnarray}
Extremization of the action leads to the vanishing of the constraints $C_1 $
and $C_2^i.$

\subsection{Coordinate reparameterization invariances}

Spatial diffeomorphisms transform $h_{ij}$ into $h_{ij} + \partial_i \xi_j+ 
\partial_j \xi_i.$ On the other hand time dependent coordinate 
transformations are as usual generated by the primary constraints which have 
been solved here. The space dependent time reparameterization 
invariance must be generated by the hamiltonian constraint.
The scalar and vector constraints generate canonically the $\xi$ and $\xi_i$
gauge variations (reparameterizations) of an observable $\phi$,  
 $\delta\phi = \{\xi C_1 -\xi_i C_2^i,\phi\},$ and
in particular :
\begin{equation}
\left\{\begin{array}{rcl} \delta h_{ij} & = & 
 \partial_i \xi_j + \partial_j \xi_i -2 Kf^2 \delta_{ij} \xi \\
 \delta p_{ij} &=& f^{-1}(-\partial_i\partial_j \xi + \delta_{ij} \Delta 
\xi ) + 2Kf^{-1} (\partial_i\xi_j + \partial_j \xi_i - 
\delta_{ij} \partial_k \xi^k) + 2K^2 f \delta_{ij}\xi
\end{array}\right.
\end{equation}

\section{Resolution of the constraints} 

\subsection{Resolution of  constraints $C_2^i$}

The vector constraint reads
\begin{equation}
\partial_j\left[p^{ij} - Kf^{-1} 
\left(2h^{ij}-\delta^{ij}h\right) \right]=0
\end{equation}
equivalently $\partial_j a^{ij}=0$
where $a^{ij}\equiv p^{ij} - Kf^{-1}\left(2h^{ij}-\delta^{ij}h\right) $
is a symmetric tensor. Hence
\begin{equation}
a^{ij}=f^{-1}\epsilon^{ikl}\epsilon^{jmn}\partial_k \partial_m P_{ln}
\end{equation}
where the prepotential $P_{ij}$ is defined up to an ambiguity 
$\delta P_{ij}=\partial_i \alpha_j+\partial_j \alpha_i.$ 
The factor $f$ is introduced here for later convenience as it simplifies the 
equations and  depends only on time.
\noindent Finally we obtain 
 \begin{equation}
\label{p}
fp^{ij}= 
\delta^{ij}(\Delta P-\partial^k\partial^l P_{kl}) + 
\partial^i\partial^k{P^j}_k + \partial^j\partial^k{P^i}_k - \partial^i 
\partial^j P - \Delta P^{ij} +  K (2h^{ij}-\delta^{ij}h ).
\end{equation}
 
\subsection{Resolution of  constraint $C_1$}
  
We deduce from above that $fp=\Delta P - 
\partial^i\partial^jP_{ij} -Kh.$
The scalar constraint can be rewritten: 
\begin{equation}
\partial_i \partial_j b^{ij} - \Delta b = 0,
\end{equation}

\noindent where  $b_{ij}=h_{ij}+2Kf^2\, P_{ij}.$
Its general solution is
\begin{equation}
b_{ij}=f\left(\epsilon_{ikl}\partial^k {\Phi^l}_j +
\epsilon_{jkl}\partial^k {\Phi^l}_i\right) +
\partial_i u_j + \partial_j u_i,
\end{equation}

\noindent with the second prepotential $\Phi$ a symmetric tensor defined up to 
the arbitrariness
$\delta \Phi_{ij}=\partial_i \beta_j + \partial_j \beta_i + \delta_{ij}
\eta$, provided one varies the third prepotential as follows
$\delta u_i = -\epsilon_{ijk}\partial^j \beta^k.$ Again a   factor $f$ has been
introduced for future simplifications.
It follows that
\begin{equation}
\label{h}
h_{ij} = f\left(\epsilon_{ikl}\partial^k 
{\Phi^l}_j + \epsilon_{jkl}\partial^k {\Phi^l}_i\right)\partial_i u_j + 
\partial_j u_i         -2Kf^2\, P_{ij} .
\end{equation}

\subsection{Gauge transformation rules and preinvariances of the prepotentials}
The coordinate reparameterization invariance must be implemented on the
prepotentials we just defined in (\ref{p}) and (\ref{h})  by some gauge
transformation laws  $(\delta P,\delta\Phi,\delta u)$
up to possible preinvariances that do not affect the original gravity fields
and momenta. We obtain first the equation:
\begin{equation}
\epsilon^{ikl}\epsilon^{jmn}\partial_k \partial_m \delta P_{ln} =
(-\partial^i \partial^j \xi + \delta^{ij}\Delta \xi)
\end{equation}
whose general solution is:
\begin{equation}
\delta P_{ij} = \delta_{ij} \xi + \partial_i \alpha_j + \partial_j 
\alpha_i.
\label{TRP}
\end{equation}

We now use  this to determine $\delta \Phi$ :

\begin{equation}
 f\left( \epsilon_{ikl}\partial^k {\delta \Phi^l}_j + \epsilon_{jkl}\partial^k 
{\delta \Phi^l}_i\right) =
\partial_i (\xi_j -u_j + 2Kf^2\, \alpha_j) +
\partial_j (\xi_i -u_i + 2Kf^2\, \alpha_i).
\end{equation}

Its general solution is given by :

\begin{eqnarray}
\delta\Phi_{ij} &=& \partial_i \beta_j + \partial_j \beta_i + \delta_{ij} 
\eta\\
\delta u_i &=& \xi_i -f\epsilon_{ijk}\partial^j \beta^k + 2Kf^2\, \alpha_i
\label{TRUPHI}
\end{eqnarray}

Preinvariances are parameterized by $\alpha_i$, $\beta_i$ and
$\eta$, they leave invariant $h_{ij}$ and $p_{ij}$, whereas gauge invariances 
-parameterized by $\xi_i$ and $\xi$- act on
these fields but leave the action invariant.

\section{Duality rotations on the prepotentials}

The benefit of having solved the constraints is that the lapse and shift 
(Lagrange multipliers $n$ and $n_i$) drop out. The new lagrangian is
  $L=p^{ij}\dot h_{ij} - H.$ We will need to reexpress it in terms of the 
prepotentials but let us start with the computation of the ``electric'' and
``magnetic'' gravitational fields (on shell). Namely let us define the 3D
tensors 
 ${\mathcal E}_{ij} \equiv Z_{i0j0}$ and 
${\mathcal B}_{ij} \equiv \tilde Z_{i0j0}.$
In the absence of torsion and on shell the remaining components of the 
Riemannian curvature are expressible in terms of these. 

\subsection{Physical fields on shell}

In the linear approximation one finds for ${\mathcal E}$
\begin{eqnarray}
-{\mathcal E}_{ij}&=&\frac{1}{2}(\partial_i \partial_j h_{00} + 
\partial_0 \partial_0 h_{ij} - \partial_i \partial_0 h_{j0} - \partial_j \partial_0 h_{i0}) \nonumber\\
 &&- K\,\partial_0 \left( h_{ij} - \frac{1}{2}f^2 h_{00} 
\delta_{ij}\right)
 \end{eqnarray}
The hamiltonian equations of motion read
\begin{eqnarray}
\dot h_{ij}&=& 2fp_{ij}-fp\delta_{ij}+ (\partial_i n_j + \partial_j n_i)
+K \, (-2h_{ij} +h\delta_{ij} + 2f^2 n \delta_{ij}) \\
-\dot p_{ij} &=& \frac{1}{2} f^{-3} \left( -\Delta h_{ij} - 
\partial_i \partial_j h + \partial_i \partial^k h_{jk} + \partial_j \partial^k 
h_{ik} + \Delta h \delta_{ij} - \partial_k\partial_l h^{kl} \delta_{ij} 
\right) \nonumber\\
&&- f^{-1}\partial_i \partial_j n + f^{-1} \Delta n \delta_{ij}
+ 2K^2f^{-1} h_{ij} + 2K^2 f n \delta_{ij} \nonumber\\
&&+K\, \left[(-2p_{ij} +p \delta_{ij}) 
- 2f^{-1} (\partial_i n_j + \partial_j n_i) 
+ 2f^{-1}\partial_k n^k \delta_{ij}\right]
\end{eqnarray}
Using $h_{00}=-2n$ and $h_{i0}=n_i$
and the scalar constraint we may check that
\begin{eqnarray}
-{\mathcal E}_{ij} &=& \frac{1}{2}f^{-2}(\Delta h_{ij} + \partial_i 
\partial_j h - \partial_i \partial^k h_{jk} - \partial_j \partial^k 
h_{ik}) \nonumber\\
&& - fK\, (p_{ij} - p\delta_{ij}) + 2K^2\, h_{ij}-\frac{K^2}{2} \,h\delta_{ij}. 
\end{eqnarray}
In terms of prepotentials and for solutions of the equations of motion
\begin{equation}
-{\mathcal E}_{ij}
= \frac{1}{2}f^{-1}\left(
\epsilon_{ikl} \partial_j \partial_m \partial^k \Phi^{lm}
+ \epsilon_{jkl} \partial_i \partial_m \partial^k \Phi^{lm}
- \epsilon_{ikl} \partial_k \Delta \Phi^l_j
- \epsilon_{jkl} \partial_k \Delta \Phi^l_i \right)
\end{equation}
\noindent Let us note that this expression does not depend explicitly on the
cosmological constant, please observe that it depends only on $\Phi$ and 
that it is gauge and
pregauge invariant. Also up to a rescaling it is the same expression as around
flat space. This  suggests that it should follow simply from the analysis of
the Weyl tensor because on shell $C_{\mu\nu\rho\sigma}=Z_{\mu\nu\rho\sigma}$ 
and of course de Sitter is conformally flat.

Let us now compute the magnetic expression ${\mathcal B} .$ By a tedious 
computation one may rewrite
\begin{equation}
-{\mathcal B}_{ij}=\frac{1}{2}f^{-1}{\epsilon_i}^{kl} {R_{klj}}^0 - 
\frac{1}{2} f^{-1} {\epsilon_i}^{kl} {Z_{klj}}^m n_m,
\end{equation}
\noindent where $\epsilon$ is the antisymetric tensor of 3D euclidean space. 
To first order
\begin{equation}
-{R_{ijk}}^0=\partial_0 \partial_{[i}h_{j]k} - 
2K\,\partial_{[i} h_{j]k} + \partial_k \partial_{[j}n_{i]} 
+ 2f^2 K\, \delta_{k[i}\partial_{j]} n.
 \end{equation}
\noindent Now to zeroth order, ${Z_{klj}}^m=0.$ Again on shell one finds
\begin{equation}
-{\mathcal B}_{ij} = {\epsilon_i}^{kl} \partial_k \left[ p_{jl} - 
\frac{1}{2} p \delta_{jl} - 2 K f^{-1} (h_{jl}-\frac{1}{4} 
h \delta_{jl}) \right].
\end{equation}
Let us now check that ${\mathcal B}_{ij}$ is symmetric in its two indices 
by contraction with the tensor $\epsilon^{ijm}$, we find that its
antisymmetric part:
\begin{equation}
-\epsilon^{ijm} {\mathcal B}_{ij} = \partial_i p^{im} - 2 f^{-1} 
K\, \partial_i h^{im} + f^{-1} K\, 
\partial^m h = 0
\end{equation}
vanishes thanks to the vector constraint. We finally discover that
      \begin{equation}
-{\mathcal B}_{ij} = \frac{1}{2}\partial_k \left[ {\epsilon_i}^{kl} (p_{jl} 
- 2 K\,f^{-1} h_{jl} ) + 
{\epsilon_j}^{kl} (p_{il} 
- 2 K\,f^{-1} h_{il} ) \right]
\end{equation}
or as a function of the superpotentials :
\begin{equation}
-{\mathcal B}_{ij}
= \frac{1}{2} f^{-1}\left(
\epsilon_{ikl} \partial_j \partial_m \partial^k P^{lm}
+ \epsilon_{jkl} \partial_i \partial_m \partial^k P^{lm}
- \epsilon_{ikl} \partial_k \Delta P^l_j
- \epsilon_{jkl} \partial_k \Delta P^l_i \right)
\end{equation}
Now this expression is proportional to the flat space case expression
of \cite{HT} which depends only on $P$ and is gauge invariant. 

\subsection{Duality and gauge invariance}

The infinitesimal duality transformation rotates ${\mathcal E}$ into 
${\mathcal B} $:
$\delta {\mathcal E}={\mathcal B}$, $\delta {\mathcal B}=-{\mathcal E}.$ 
Similarly $\delta \Phi = P$, $\delta P=-\Phi.$
On the other hand if one applies a $\pi/2$ duality rotation one obtains  
$\Phi' =  P$ and $ P' =  -\Phi.$  One may define a dual metric $h'_{ij}$
given in terms of prime variables as in (\ref{h}) with due replacements.
As a result we have at our disposal (in this approximation) a one parameter
family of metrics on 4D space provided the  action  of the duality
transformation is non trivial. It is an  important question 
to determine whether the 4-manifolds can be
assembled into a higher dimensional structure for instance. 

But let  us focus on  discrete duality for the time being. Is there a dual 
coordinate reparameterization invariance? On the face of it there is a single
spacetime on which the metrics are computed. Actually in the flat space 
perturbative  analysis of \cite{HT} the prepotential $u$ was
discarded by a coordinate choice (essentially a spatially harmonic  
gauge if one fixes also appropriately the pregauge invariances). But we could 
try and preserve the coordinate reparameterization invariance and  make it
compatible with duality. This means probably that one should introduce 
a dual ``spatial'' coordinate reparameterization and  that means in
turn that one should presumably have a second set of coordinates, something
string theorists might appreciate.  

We may collect the set of transformation
rules  with dual pairs of parameters to emphasize their
symmetry under duality:
\begin{eqnarray}
\delta P_{ij} &=& \delta_{ij} \xi + \partial_i \alpha_j + \partial_j 
\alpha_i.\\
\delta\Phi_{ij} &=& \delta_{ij} \eta + \partial_i \beta_j + \partial_j \beta_i\\
\delta u_i &=& \xi_i -f\epsilon_{ijk}\partial^j \beta^k + 2Kf^2\, \alpha_i
\label{deltas}
\end{eqnarray}
We may now ask for the fate  of the gauge and
pregauge invariances under duality. The absence of a dual $v_m$ to $u_m$ 
breaks the duality symmetry, however $u$ is pure diffeomorphism gauge and 
we may ignore it for the rest of this paper if we ignore $v.$
Introducing primes on the  parameters as above after discrete duality we get
for the dual parameters:  
\begin{eqnarray}
\alpha'_i &=& -\beta_i\\
\beta'_i &=&  \alpha_i\\
\xi' &=& - \eta\\
\eta' &=& \xi 
\end{eqnarray}
It maybe worthwhile to investigate space(time) doubling as suggested above,
we would need 
$$\xi_i' = \delta u'_i + f \epsilon_{ijk} \partial^j \alpha^k + 2Kf^2\,
\beta_i.$$

\section{Invariance of the action} 

We are now ready to rewrite the action in terms of $P$,$\Phi$ ($u$ disappears
because of coordinate reparameterization invariance).

\subsection{Action in terms of prepotentials}

We saw that after solving the constraints the lagrangian simplifies to  
$L=p^{ij}\dot h_{ij} - H.$  Let us call $H_1$ the part of $H$ that depends on 
$\partial h$ and $H_2$ the rest of it with quadratic terms $p^2$, $ph$, $h^2.$ 
Let us define ${\mathcal P}^{ij} \equiv f p^{ij} - K (2h^{ij}-\delta^{ij}h) $ :
\begin{equation}
fH_2 ={\mathcal P}^{ij}{\mathcal P}_{ij} - 
\frac{1}{2}(\Delta P - \partial_i \partial_j P^{ij})^2
+ 2K\, h_{ij}{\mathcal P}^{ij} + K^2\left(h^{ij}h_{ij} - \frac{1}{2}h^2\right)
\end{equation}
Let us now compute $p^{ij}\dot h_{ij}$ , it is equal up to a total derivative 
to:
\begin{equation}
fp^{ij}\dot h_{ij} = K^2\left(h^{ij} h_{ij} - 
\frac{1}{2}h^2\right) +{\mathcal P}^{ij}\dot h_{ij}.
\end{equation}
Hence
\begin{equation}
f L= {\mathcal P}^{ij}\dot h_{ij} -
2K{\mathcal P}^{ij} h_{ij} - \left[{\mathcal
P}^{ij}{\mathcal P}_{ij} - \frac{1}{2}(\Delta P - \partial_i \partial_j
P^{ij})^2\right] - fH_1
 \end{equation}
We obtain term by term up to total derivative terms again: 
\begin{eqnarray}
{\mathcal P}^{ij}{\mathcal P}_{ij} &-& \frac{1}{2}(\Delta P - \partial_i 
\partial_j P^{ij})^2
\nonumber \\
 &=& \Delta P^{ij} \Delta P_{ij} - \frac{1}{2} \Delta P 
\Delta P + 2 \partial_i P^{ij} \Delta \partial^k P_{jk} + \Delta P 
\partial_i \partial_j P^{ij} + \frac{1}{2}(\partial_i \partial_j P^{ij})^2\nonumber
\end{eqnarray}

\begin{eqnarray}
{\mathcal P}^{ij} \dot h_{ij} &=& 2f\epsilon_{ijk}\partial^j 
\left[\dot \Phi^k_l+K\Phi^k_l\right] 
(\partial^l \partial_m P^{mi} - \Delta P^{il}) - 3K^2f^2 ( P\Delta P  
- P_{ij}\Delta P^{ij}
\nonumber\\
&&-2 P \partial_i \partial_j P^{ij} -2 \partial_i P^{ij} \partial^k P_{jk})
\end{eqnarray}

\begin{eqnarray}
2K\,{\mathcal P}^{ij}  h_{ij} &=& 
4Kf\,\epsilon_{ijk}\partial^j \Phi^k_l 
(\partial^l \partial_m P^{mi} - \Delta P^{il}) - 4K^2f^2
( P\Delta P -  P_{ij}\Delta P^{ij}
\nonumber\\
&& -2P \partial_i \partial_j 
P^{ij} -2 \partial_i P^{ij} \partial^k P_{jk})
\end{eqnarray}

\begin{eqnarray}
H_1 &=& f^{-1} (\Delta \Phi^{ij} \Delta \Phi_{ij} - \frac{1}{2} \Delta 
\Phi \Delta \Phi + 2 \partial_i \Phi^{ij} \Delta \partial^k \Phi_{jk} + 
\Delta \Phi \partial_i \partial_j \Phi^{ij} + \frac{1}{2}(\partial_i 
\partial_j \Phi^{ij})^2) \nonumber\\
 && + K^2f (P\Delta P - P_{ij} \Delta P^{ij} - 2P 
\partial_i \partial_j P^{ij} - 2 \partial_i P^{ij} \partial^k P_{jk})
 \nonumber\\
 &&- 2K\,\epsilon_{ijk}\partial^j \Phi^k_l 
(\partial^l \partial_m P^{mi} - \Delta P^{il})
\end{eqnarray}

At last :

\begin{eqnarray}
L &=& -f^{-1} \left( \Delta \Phi^{ij} \Delta \Phi_{ij} - \frac{1}{2} 
\Delta \Phi \Delta \Phi + 2 \partial_i \Phi^{ij} \Delta \partial^k 
\Phi_{jk} + \Delta \Phi \partial_i \partial_j \Phi^{ij} + 
\frac{1}{2}(\partial_i \partial_j \Phi^{ij})^2 \right) \nonumber\\
 &&
 - f^{-1} \left(\Delta P^{ij} \Delta P_{ij} - \frac{1}{2} \Delta P 
\Delta P + 2 \partial_i P^{ij} \Delta \partial^k P_{jk} + \Delta P 
\partial_i \partial_j P^{ij} + \frac{1}{2}(\partial_i \partial_j 
P^{ij})^2\right) \nonumber\\
&&
 + 2\epsilon_{ijk}\partial_l \dot \Phi^{kl}  \partial^j \partial_m P^{im}
- 2\epsilon_{ijk}\partial^j \dot \Phi^{kl} \Delta {P^i}_l
\end{eqnarray}

The first two lines are manifestly invariant under duality
$\delta \Phi =  P, \delta P = - \Phi.$
Let us vary the penultimate term
$A \equiv 
\epsilon_{ijk}\partial_l \dot \Phi^{kl} \partial^j \partial_m P^{im}$ :

\begin{eqnarray}
\delta A = 
\epsilon_{ijk}\partial_l \dot P^{kl} \partial^j \partial_m P^{im}
 \nonumber 
 - \epsilon_{ijk}\partial_l \dot \Phi^{kl} \partial^j \partial_m \Phi^{im}
\end{eqnarray}

Using antisymmetry in $i$ and $k$ and up to a spatial derivative term :

\begin{equation}
\delta A = \frac{1}{2}\partial_0\left( \epsilon_{ijk}\partial_l P^{kl} 
\partial^j \partial_m P^{im} - \epsilon_{ijk}\partial_l \Phi^{kl} 
\partial^j \partial_m \Phi^{im}\right)
\end{equation}
The last term in the lagrangian is also a time derivative and the full 
action is therefore duality invariant.

\subsection{Manifestly dual expression}

We have extended the noncovariant action of \cite{HT} to the de Sitter background with
minor modifications by powers of $f.$ We leave it for another project to
render the whole presentation Lorentz covariant. But let us conclude here with
a manifestly dually symmetric action. 
Defining the doubled prepotentials
\begin{equation}(A^{mn}_a)_{a=1,2}=(P^{mn}, \Phi^{mn})\; , \end{equation}
we get
\begin{eqnarray}
L&=&\epsilon^{ab}\epsilon^{mnp}
(\partial^q\partial_n\partial_rA_{aqp}-\Delta
\partial_nA_{arp})\dot{A}^r_{bm} \nonumber \\
&&-\delta^{ab}f^{-1}\left\{\Delta A_{amn}\Delta
A_b^{mn}+\frac{1}{2} \partial^m\partial^n
A_{amn}\partial^p\partial^{q}A_{bpq} + \partial^m\partial^n
A_{amn}\Delta
A_b \right.\nonumber\\
&&\left.- 2 \partial_m \partial_n A_a^{mp} \partial^n \partial^q A_{bpq}
- \frac{1}{2} \Delta A_a \Delta A_b \right\} \;. \end{eqnarray}

The Noether charge that generates the duality rotation is strictly identical
to that of \cite{HT}. One may extend duality  also outside the
constraint surface by the same general procedure too.

\section{Conclusion}

Let us hope that the $\Lambda$-instantons have more uses - they deserve closer
investigation -, and that the $Z_{\mu\nu\rho\sigma}$
tensor finds its truly geometrical interpretation. 
We have shown that duality invariance of quasi-flat gravity can be extended
to  the quasi-de Sitter case, which certainly is a large perturbation away.
It is an encouraging  indication that the duality features of 4D General
Relativity may be simpler than might be 
expected from the dualization procedure of 
Eddington and Schroedinger \cite{ES,Jprogress}.  

\acknowledgments
We benefited from discussions with Y. Dolivet, M. Henneaux and J. Iliopoulos
and U. Lindstrom.
------------------------------------------------------------------------


\begin{thebibliography}{999}
\bibitem{Jprogress} B. Julia, M. Perry and I. Dolgachev work in progress 
announced at the Lebedev International conference on Theoretical 
Physics, Moscow April 15, 2005.
\bibitem{HT} M. Henneaux and C. Teitelboim, Phys.Rev. D71 (2005) 024018. 
\bibitem{JU} In his talks on the subject B.J. usually emphasizes the 
challenge to ``understand'' a geometrical reason for those hidden
dualities that do not come obviously from dimensional reduction. In the Carg\`
ese lectures hep-th/9805083 he argued for the fusion of the spacetime 
symmetries and the dualities inside
a group that is independent on the dimension of spacetime under consideration.
The reduction of this group into the above two factors is done differently in 
each dimension. This is one example of action of the duality groups above their
observed dimension.   
\bibitem{BB}A.A. Belavin and D.E. Burlankov, Phys.Lett.58A (1976) 7. See also
M.F. Atiyah, N.J. Hitchin and I.M. Singer, Proc.Roy.Soc.London A362 (1978) 425. 
\bibitem{MDM}S.W. MacDowell and F. Mansouri, Phys.Rev.Lett. 38 (1977) 739, erratum 
ibid. 38 (1977) 1376.
\bibitem{DS}S. Deser and D. Seminara, Phys.Lett.B607 (2005) 317. 
\bibitem{EF}T. Eguchi and P.G.O. Freund, Phys.Rev.Lett.37 (1976) 1251.
\bibitem{DN}S. Deser and R. Nepomechie, Phys.Lett.97A (1983) 329.
\bibitem{HA} P.A.M. Dirac, Proc.Roy.Soc.A246 (1958) 333,\\
R. Arnowitt, S. Deser and C.W. Misner (1962) 
reproduced in the archive gr-qc/0405109 \\
and L. Abbott and S. Deser Nuc.Phys.B195 (1982) 76.
\bibitem{ES} E. Schroedinger, Proc.Roy.Ir.Ac. 51 (1947) 163.



\end{thebibliography}
\end{document}